**Exploring the Drivers and Barriers to Uptake for Digital Contact Tracing**
*Andrew Tzer-Yeu Chen, Kimberly Thio*
*Koi Tū: The Centre for Informed Futures, The University of Auckland, New Zealand*
*Correspondence: andrew.chen@auckland.ac.nz*

**Abstract**

Digital contact tracing has been deployed as a public health intervention to help suppress the spread of Covid-19 in many jurisdictions. However, most governments have struggled with low uptake and participation rates, limiting the effectiveness of the tool. This paper characterises a number of systems developed around the world, comparing the uptake rates for systems with different technology, data architectures, and mandates. The paper then introduces the MAST framework (motivation, access, skills, and trust), adapted from the digital inclusion literature, to explore the drivers and barriers that influence people's decisions to participate or not in digital contact tracing systems. Finally, the paper discusses some suggestions for policymakers on how to influence those drivers and barriers in order to improve uptake rates. Examples from existing digital contact tracing systems are presented throughout, although more empirical experimentation is required to support more concrete conclusions on what works.

**Keywords:** digital contact tracing, covid-19, public health policy, behaviour change

## 1. Introduction

Many attempts have been made by public health authorities around the world to minimise the spread of Covid-19 [1]. Contact tracing is a fundamental method of epidemiology to find people ("contacts") who may have been potentially exposed by a person who has tested positive for an infectious disease such as Covid-19 (an "active case") [2]. Those contacts can then be ordered to isolate and separate themselves from the community, preventing them from passing the disease on further [3]. This is based on two types of information – interactions with individuals who were physically proximate to the active case, or the locations where the active case and other individuals have been in order to infer interactions. The goal of contact tracing is full coverage – to find all possibly infected people and cut off the chains of transmission. However, this can be challenging because manual processes rely on time-consuming interview processes and fallible human memory. The relatively short incubation time, the high infectiousness, and the global spread of Covid-19 make it difficult for manual contact tracing to keep the reproduction rate down.

Digital contact tracing (DCT) introduces digital technologies into this process, with the aim of improving the speed and completeness of contact tracing [4, 5, 6]. DCT can help identify and alert contacts at large scales and in a semi-automated way, mitigating human capacity limitations. It is promoted in some containment models due to its effectiveness alongside testing [7, 8], with some researchers arguing that it is imperative for containment to take effect [9, 10]. However, theoretically simple in its logic and use, the adoption of this tool has presented challenges, both ethically and in its effectiveness [11].

With a rapid response to the Covid-19 pandemic, there have been a large variety of tools adopted, depending on the technological capacity of the jurisdiction, on the capacity of the contact tracing system, on data privacy laws, and many other factors. What is common is that most jurisdictions have found it difficult to achieve a high uptake or adoption rate, limiting the potential effectiveness of this tool. As a relatively new tool, there is limited research on DCT – a rapid *Cochrane* review found that prior to the Covid-19 pandemic, there were only six cohort studies and six modelling studies about the use of DCT, all in relatively constrained settings [12], so there is little evidence for policymakers to draw upon when thinking about how to best use DCT. This paper seeks to conceptually explore the underlying drivers and barriers for why individuals choose to participate or not in DCT through a literature review approach, and delves into some policy suggestions for how to potentially influence those drivers and barriers in order to improve uptake rates. While the introduction of vaccines may mark a transition to a new phase of the Covid-19 pandemic, there may be lessons valuable for the use of DCT in managing more virulent strains of Covid-19, or in future pandemics.

## 2. Characteristics of Digital Contact Tracing

We separate the different approaches based on three primary characteristics: the technology used, whether the system architecture is centralised or decentralised, and how mandatory the use of the tool is. This Section highlights the diversity of approaches adopted in various jurisdictions around the world, leading to differences in effectiveness. These primary characteristics can then influence second-order characteristics, such as ease of integration with public health systems, privacy, and equity. These characteristics may then influence individuals' choices on whether to participate in DCT or not, which are discussed in Section 3 and 4.

### 2.1 Technology

The three key parts of the technology used in DCT are the sensor technology (i.e. Bluetooth, GPS, QR Codes via Cameras), the system hardware (i.e. smartphones, wearable devices like smart watches or bracelets), and the software protocols that control and drive the tool (including any algorithms used to interpret data and find contact matches). The most common hardware approach globally relies on smartphones, with Singapore leading the use of wearable devices [13]. There is more variation in the sensor technology, with some systems using only one sensing technology, while there is a trend towards combining multiple approaches in the same tool (e.g. Bluetooth-based approaches as well as QR code scanning in the same app) [14].

Bluetooth-only DCT apps are the most common approach, such as in France (TousAntiCovid), Australia (COVIDSafe App), and Germany (CoronaWarn). Bluetooth technology allows smartphones and other compatible devices to keep a record of other devices that have been in physical proximity, with an assumption that those devices are carried by individuals who have therefore been in physical proximity with another person [15]. Therefore, when a person tests positive for Covid-19, this list of devices can be used to identify that person's contacts, and they can be notified of the necessity to self-isolate and/or take a test [16]. Throughout this paper, we interchangeably refer to people and devices as capable of being a "contact".

Bluetooth also offers compatibility with a variety of hardware technologies. Although smartphones are the most common, Singapore also has a wearable token that can be easily carried by individuals without requiring any interaction with an app or smartphone system [17]. These tokens were deliberately made for people without smartphones or without the skills/confidence to operate those smartphones, especially the more vulnerable elderly population. The wearable tokens allow for better inclusion of individuals, increasing the uptake rate of DCT.

Alternatively, the use of GPS on top of Bluetooth allows tracing of both the people around an individual as well as the location of an individual. Although this approach is not as common as Bluetooth-only approaches, some jurisdictions such as India (Aarogya Setu) and Norway (Smittestop) use both. By using the GPS equipped on smartphones, relatively accurate location estimates can be produced, allowing the software to determine when two phones have been in physical proximity for enough time to be considered a contact [4]. The combination of these two technologies can allow for more accurate tracing, and therefore more accurate identification of contacts. Some tools also allow users to draw their paths on a map after the fact, and the system then checks that against the GPS trails recorded for active cases [18].

The use of QR codes is also gaining popularity for DCT globally. When using venue-based QR code technology, buildings and public places (including public transportation) place QR codes in visible places, such as at shopfronts or on the back of a bus seat. Individuals then use cameras on their smartphones to scan the code when they enter a location. This approach therefore provides location information for individuals, although at a less granular level than GPS. New Zealand (NZ COVID Tracer) initially used QR codes only, until December 2020 when Bluetooth was introduced into the same app [19]. QR codes are also used in Singapore's SafeEntry app, which is intended to be used in conjunction with their TraceTogether Bluetooth-based system. More recently, each State in Australia has set up their own venue-based QR code system that operates alongside the Federal Bluetooth-based CovidSafe app [20], and the United Kingdom adopted and adapted New Zealand's QR code approach alongside Bluetooth [21].

QR Codes are also used in China, except that instead of codes being assigned to venues, they are assigned to individuals. These are then scanned by guards at health checkpoints, which queries a central database and returns a message indicating whether the user is healthy (green), required to stay at home for 7 days (yellow), or needs to be quarantined (red) [22]. There is limited public information on how these colours are assigned to individuals, but it has been reported that user location is one component used [23]. Israel has also adopted a similar approach for vaccination certificates, although there are concerns of wide-spread forgery [24].

### 2.2 Data Architecture

A key characteristic is where the data in a DCT system should be stored. If a system is centralised, then the user's data will be generated, processed, and stored in a central server maintained by public health authorities [25]. In a decentralised approach, the data is generated, processed, and stored on each user's own device, and only shared with public health officials if necessary [26]. Centralised approaches have led to negative public opinion due to privacy concerns stemming from the limits of trust in government. However, some public health officials have argued that a centralised approach is necessary to allow access to information critical for contact tracing [27].

It has been argued that privacy can be protected in a centralised system by anonymising users through the assignment of a unique ID code before their data is sent to the central database [14]. Legislative protections on data use, such as those seen in Australia, can also help provide confidence that the data collected will not be misused for other purposes [28]. Privacy advocates tend to prefer a decentralised approach, because governments cannot easily access a user's list of contacts without their knowledge or permission. Instead, public health authorities issue exposure notifications to each device, containing information about where and when potential exposures occurred, or information about which ID numbers correspond to individuals who were recently infectious, which are then compared against logs on the device itself [14].

Systems that use a centralised approach include ROBERT, PEPP-PT, and BlueTrace (in its initial approach) [25]. Some jurisdictions, such as Switzerland, initially backed a centralised approach like PEPP-PT, but then subsequently shifted to a decentralised approach (DP-3T) to help protect privacy [29]. Other examples of decentralised systems include TCN and the Google-Apple Exposure Notification Framework (ENF), which is now the most commonly adopted protocol around the world [25]. The Exposure Notification Framework takes advantage of Google and Apple's control of Android and iOS operating systems (which collectively account for 99% of the world's smartphones) to lock down data on the device and prevent other apps from accessing Bluetooth contact logs.

### 2.3 Compulsion

Lastly, whether the DCT tool is made mandatory or voluntary by a government is another key differentiating characteristic. Most jurisdictions have opted to make the app voluntary, relying on the goodwill and initiative of the citizens, but a minority of jurisdictions have requirements for some or all people to participate in DCT. In Turkey (Hayat Eve Sığar), it was initially only mandatory for those who tested positive to use their app, but it is now mandatory for all of its citizens [30]. India officially says their app (Aarogya Setu) is voluntary, but participation is mandatory for citizens living in containment zones and for all government and private sector employees, effectively making it mandatory for most people [31]. China (Alipay Health Code) claims that they have not made their app mandatory, but it is required for access to enter many buildings and public areas, making it effectively mandatory if people want to participate in everyday society [23]. There have also been discussions of quasi-mandates or unenforced mandates in some jurisdictions, where a government makes participation in DCT compulsory, but does not exercise any enforcement action against individuals who do not comply. The Google-Apple Exposure Notification Framework prohibits governments that use their approach from making participation mandatory, and requires that users be given a genuine opt-in choice [32].

### 2.4 Comparative Analysis of Uptake Rates

To help understand the impact of these three characteristics on the adoption of DCT tools, we compiled a table of data from publicly available sources. In Table 1, we summarise tools from 14 jurisdictions, listing the sensor technology, hardware, and software protocols, whether the tool is centralised or decentralised, whether the tool is mandatory or not, and an approximate uptake rate. The uptake rate can be very difficult to determine as there are many ways to interpret it. For example, do we include all people in the jurisdiction, or only adults? Do we include all adults, or only those who have smartphones? If there are multiple tools, will there be some people using more than one, or can we sum the participation numbers together? Are there official statistics available from health agencies, or are we using less reliable secondary sources? Because of these doubts, we provide an approximate range for the uptake rate in each jurisdiction, based on desk research in December 2020 - February 2021. In most cases, we only have data on the total number of downloads, which is not necessarily the same as the number of people who are actively using the tool and participating on a daily basis.

### 2.5 The Importance of Uptake Rate

DCT is best viewed as an augmentation of manual-led contact tracing, rather than a completely automated replacement. Scientific models have suggested how much of the population might need to participate in a DCT system to collect enough information to be effective in reducing the reproduction rate of Covid-19. One of the first pieces of modelling work, by Ferretti et al, suggested that, if those over 70 were also shielded or protected by good preventative measures, then a 56% uptake rate would be sufficient to bring the reproduction rate below 1 and suppress the pandemic [10]. It is commonly misreported from this paper that a 60% uptake rate is therefore required for DCT tools to be effective, when the model shows that the protective effects could exist at every level of uptake, in that it would still reduce the reproduction rate but could have less of an impact [33, 34]. Alternatively, one model in New Zealand recommends a higher uptake rate of at least 75%, with 90% of contacts recorded through digital means, and accompanied by human-led contact tracing of an active case's contacts [35].

The Ferretti paper also suggests that DCT would need to be implemented alongside social/physical distancing interventions [36]. Other models show that a combination of interventions, such as with quarantine and isolation, is required to reduce the spread of Covid-19, rather than solely relying on DCT [37]. However, these models show that overall, the uptake rate is the strongest determinant of the effectiveness of DCT in reducing the reproduction rate. There is a notable distinction between contexts where the aim is to suppress the number of daily cases below an acceptable level, and contexts where the aim is elimination. A number of models suggest that for small outbreaks or scenarios with a low number of cases, digital technology plays a larger role by improving the speed of contact tracing and therefore cutting off chains of transmission faster [35, 37]. In that context, a high uptake rate can allow a government to have more confidence in loosening lockdown restrictions and relaxing social/physical distancing measures [38], predicated on the assumption that contact tracing leads to effective isolation of infectious individuals. DCT may just be one tool in pandemic response, but a high uptake rate could have a large contribution towards managing the spread of Covid-19 on a medium-to-long-term basis [8].

| Jurisdictions (DCT Tool) | Technology | Hardware | Protocols | Centralised / Decentralised | Mandatory / Voluntary | Approximate Uptake Rate |
|---|---|---|---|---|---|---|
| **Qatar (Ehteraz)** | Bluetooth + GPS | Smartphones | Custom-made | Centralised | Mandatory (requires access to photos + must download or face prison for 3 years or max $55,000 fine) | 85-95% |
| **China (The Alipay Health Code)** | QR code | Smartphones | Alipay & Tencent | Unsure | Officially voluntary, effectively mandatory (required to access most public buildings or transport) | 75-85% |
| **Singapore (TraceTogether and SafeEntry)** | Bluetooth + GPS + QR codes | Smartphones + Wearable tokens | BlueTrace | Decentralised Bluetooth, Centralised QR code scans | Generally voluntary, mandatory for people attending high risk activities or large events since Dec 2020 | 65-75% |
| **New Zealand (NZ Covid Tracer)** | QR Codes, with Bluetooth introduced in Dec 2020 | Smartphones | Google/Apple API | Decentralised | Voluntary | 55-65% |
| **Ireland (COVID Tracker App)** | Bluetooth | Smartphones | Google/Apple API | Decentralised | Voluntary | 45-55% |
| **Norway (Smittestopp)** | Bluetooth + GPS | Smartphones | Google/Apple API | Decentralised | Voluntary | 45-55% |
| **Iceland (Rakning C-19)** | GPS | Smartphones | Sensa | Decentralised | Voluntary | 35-45% |
| **Australia (COVIDSafe App)** | Bluetooth | Smartphones | Initially BlueTrace, later Herald | Decentralised, State QR code apps are Centralised | Voluntary | 25-35% |
| **Switzerland (SwissCovid)** | Bluetooth | Smartphones | Initially DP-3T, now Google/Apple API | Decentralised | Voluntary | 25-35% |
| **Germany (CoronaWarn)** | Bluetooth | Smartphones | Google/Apple API | Decentralised | Voluntary | 25-35% |
| **United Kingdom (NHS COVID-19)** | Bluetooth + QR Codes | Smartphones | Google/Apple API | Decentralised | Voluntary | 25-35% |
| **Uruguay (CoronavirusUy)** | Bluetooth | Smartphones | Google/Apple API | Decentralised | Voluntary | 15-25% |
| **India (Aarogya Setu)** | Bluetooth + GPS | Smartphones | The Aarogya Setu Data Access and Knowledge Sharing Protocol | Centralised | Initially voluntary, effectively mandatory from April/May 2020 onwards | 15-25% |
| **France (TousAntiCovid)** | Bluetooth | Smartphones | ROBERT | Centralised | Voluntary | 5-15% |
| **South Africa (COVID Alert SA)** | Bluetooth | Smartphones | Google/Apple API | Decentralised | Voluntary | 0-10% |

Table 1. Digital contact tracing tools in selected jurisdictions, comparing the approximate uptake rate achieved across different system characteristics, as of December 2020 - February 2021.

Different jurisdictions have then used the modelling work to set different targets for uptake rates, with Australia aiming for 40% of the population [32], Singapore aiming for at least 70% of the population [33], and the United Kingdom aiming for 80% of smartphone owners [34]. As Table 1 begins to show, the uptake rate falls below the 60-80% targets of many jurisdictions. Surveys conducted in a number of jurisdictions have shown high levels of acceptance and willingness to participate in DCT, yet there appears to be a large behaviour-intention gap in many jurisdictions [39, 40, 41].

Some patterns show that the choices in system and policy design can have a strong influence on the uptake rate of the app. Uptake rate is generally high when DCT is made mandatory, yet most governments are reluctant due to fears of infringing on human rights and personal autonomy [42]. Some jurisdictions may also have rights-based legislative frameworks that prohibit making participation in contact tracing mandatory. This leads to questions of privacy protection – even when the tool is voluntary, if the privacy protections are ambiguous, the uptake rate will likely stay low [43, 44].

Technological issues can also have an impact on uptake rate. Concerns about error rates and the potential for false positives or false negatives with Bluetooth technology may lead to reduced confidence from potential users [16, 45, 46]. GPS-based approaches also attract concern about accuracy indoors and in multi-level buildings [4]. GPS tracks could also be de-anonymised, revealing personal information about a person's activities or daily routines [14]. As QR codes require active participation, there are concerns that users may miss scanning codes that are critical later on [47]. These considerations contribute towards perceptions of unreliability, both in terms of effectiveness and protecting privacy.

The next section explores these ideas further and presents a framework for understanding the factors that influence uptake rate by analysing the decisions that are made by individuals about whether they participate in DCT or not.

### 3. Drivers and Barriers to Uptake

While DCT is a relatively new concept, there have been a number of frameworks suggested already in the literature to help evaluate DCT tools or understand individual decisions about participation in DCT [43, 48, 49, 50, 51, 52, 53, 54]. For example, the COVIDTAS framework evaluates the data privacy and security of various DCT apps [54], while Munzert et al. explore the effectiveness of monetary incentives on DCT participation [49]. Over time, it will also be important to update these evaluations as experiences and contexts change [56]. As more frameworks are created, it is difficult to comprehensively compare against all of them. This paper does not focus on evaluating effectiveness, privacy, or data security of various tools, but seeks to understand how governments can support increased uptake of digital contact tracing tools.

In this paper we borrow a framework from a similar context, which appears to have originated from the UK Government's Digital Inclusion Strategy [57]. It views digital inclusion, a policy objective for every person to have equitable opportunity to participate in society using digital technologies, as being built on four pillars: access, capability, motivation, and trust. This framework has also been used by other governments in their digital inclusion approaches [58], having developed over time from viewing digital inclusion as an access-only problem (i.e. "just give people laptops") to a more sophisticated understanding of the challenges that face the digitally excluded. We see similarities between those challenges and encouraging uptake of DCT tools because of the digital nature of this health intervention.

For the purposes of our discussion, we re-order and slightly rename the pillars into what we have named the MAST framework: motivation, access, skills, and trust. This framing helps us consider the factors in DCT, which has to target the entire population rather than just the digitally excluded. While we use the MAST categories to structure this section, the drivers and barriers are identified for DCT and may not overlap with those for digital inclusion. Obviously, the factors do not neatly fit into only one category, and there are overlaps when considered in context. We also note that these drivers and barriers affect each individual's decision-making processes in different ways. The factors below were identified through literature review and discussion with academics and civil society experts, as well as drawing upon (unpublished) opinion polling and focus group work conducted by the New Zealand Ministry of Health on perceptions of DCT. The factors are summarised in Table 2.

| Motivation | Access |
|---|---|
| 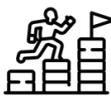 • Mandatory or Voluntary<br>• Changes in Perception of Risk<br>• Changes in Perception of Effectiveness<br>• Convenience<br>• Bandwagon Effect | 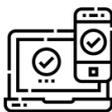 • Hardware Access<br>• Infrastructure Access (e.g. Internet, Mobile Data)<br>• Software Access + Accessibility |
| **Skills** | **Trust** |
| 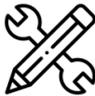 • Technology Skills and Confidence<br>• Literacy Skills | 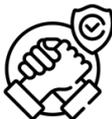 • Trust in Government<br>• Trust in Corporations<br>• Data Architectures and Privacy<br>• Security Concerns |

Table 2. Factors associated with the drivers and barriers to uptake for digital contact tracing. Icons: flaticon.com

### 3.1 Motivation

Do people want to use DCT? Are they motivated enough to spend the time and effort to participate? This category looks at reasons that move a person to be more favourable or more negative towards DCT generally, and therefore push them towards or away from participation. We identify five main factors here:

#### 3.1.1 Mandatory or Voluntary

As suggested in Section 2.4 and 2.5, whether the government mandates the use of their DCT tool or relies on voluntary adoption appears to strongly influence the uptake. When made mandatory, the fear of repercussions for not participating strongly motivates people to use the tool. For example, in Qatar, their mandatory app (Ehteraz) must be downloaded and given access to the user's photos, otherwise the individual will face prison for up to 3 years or a fine of up to QAR$55,000 [59]. People are required to show that they have the app before they can access public transport or access supermarkets. Where the use of DCT is voluntary, governments face a tougher challenge in having to motivate people through other drivers, otherwise the default action for most people is to do nothing.

#### 3.1.2 Changes in Perception of Risk

When there is a high and escalating number of cases in a jurisdiction, people may more easily perceive Covid-19 as a high-risk disease, leading to greater motivation to use their jurisdiction's DCT tool to help protect themselves. This is sometimes analogously seen in the sudden rush in downloads of emergency management apps after natural disasters [60]. The opposite effect may occur when the number of cases in a jurisdiction is low, or reducing steadily, as people feel that the risk is low and no action is required. This behaviour is well characterised in the pre-COVID psychology literature as reactance [61], and has also been studied in the Covid-19 context for face masks and vaccinations [62, 63]. A survey in Switzerland highlighted perception of personal health risk as a strong driver for acceptance of DCT [64]. Other surveys and interviews exploring the role of risk perception in uptake of DCT include those undertaken in the US [65], China [66], and Germany [67].

Perceptions are relative and context-specific – an increase in the daily case count by one has a much larger impact on the perception of risk in a jurisdiction that has had no cases for a hundred days than in a jurisdiction where daily cases average in the thousands. This has been clearly apparent in the use of New Zealand's DCT app (NZ COVID Tracer), where usage significantly increases whenever there are Covid-19 cases emerging in the community, and then when there are no cases the usage reduces over time [68], as shown in Figure 1. The effect of perception changes can also evolve over time. People can also become fatigued as the pandemic continues; hearing about COVID-19 cases and its continuous rise can have the effect of normalising the high level of risk, consequently reducing people's motivation to act and use DCT [69]. However, there are dissenting views on the role of pandemic fatigue in influencing behaviours [70, 71]. Research in Poland has also shown that the underlying ideology of individuals (e.g. acceptance of authoritarianism) influences perceptions of personal threat and risk, and therefore their acceptance of surveillance tools like DCT [72].

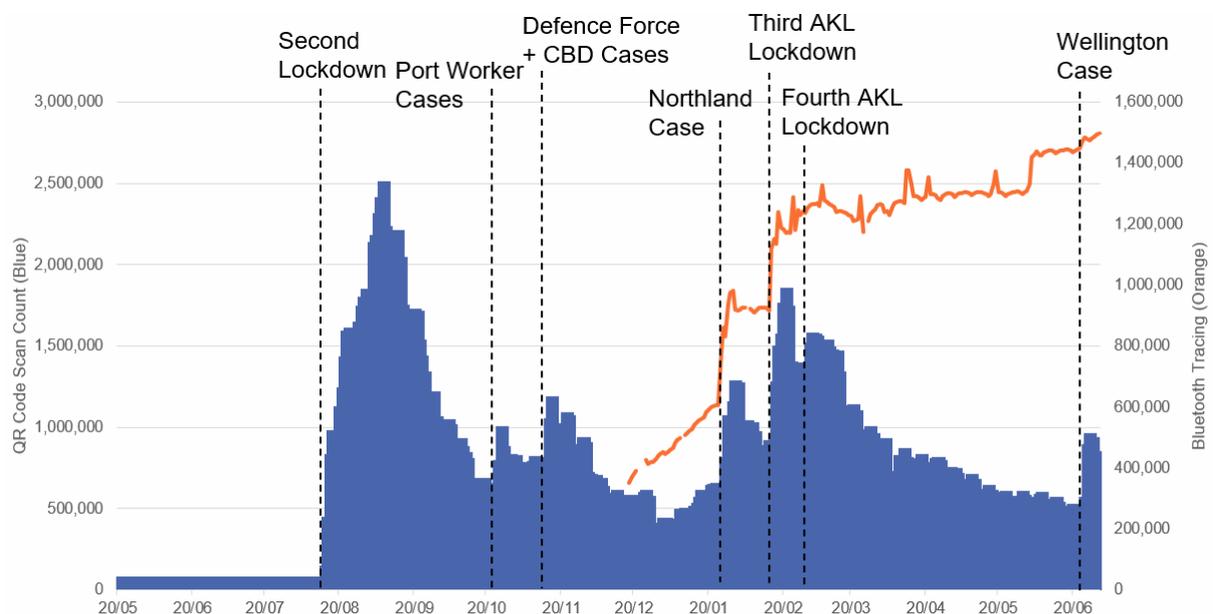

Figure 1. Graph showing usage of the NZ COVID Tracer app between May 2020 and June 2021, showing that spikes in usage (based on daily QR code scan counts in blue, or number of devices with Bluetooth Tracing enabled in orange) follow soon after active cases appear in the community.

### 3.1.3 Changes in Perception of Effectiveness

Complementary to the perception of risk or danger, people may also evaluate whether DCT is likely to be effective in mitigating that risk. If people believe that using DCT is effective, in that it can reduce the likelihood that they or their community will be infected with Covid-19, or that it could reduce the severity or duration of lockdowns, then it may increase their motivation to participate. Conversely, a belief that the tool does not contribute to public health efforts, or that the error rates are too significant, can reduce motivation to participate. Pandemic fatigue can also reduce motivation, as people may perceive an ongoing growth in cases, despite participation in DCT, as a failure of the tool to suppress the spread of the disease. In an analysis of newspaper coverage across Germany, Austria, and Switzerland, the "functional efficacy" of DCT tools was identified as a key topic of discussion, highlighting technical limitations and their effect on adoption of DCT tools [73].

### 3.1.4 Convenience

The ease of use can have a significant effect on motivation, particularly in that poor usability creates a barrier that individuals have to be motivated to overcome. In the iOS version of Singapore's TraceTogether app, the app must be constantly open in the foreground in order to use the Bluetooth functionality, which prevents users from using their phone to do other things and requires extra actions in comparison to other Bluetooth approaches [74]. Users may also worry about DCT apps draining their phone batteries more quickly or using a lot of mobile data, creating a barrier to participation even where this is only a perception and not technical reality. For QR code-based systems, if venues do not have codes available or codes are hard to find, then individuals are more likely to forget to scan – New Zealand had to counter this by making it mandatory for businesses to display QR codes [75].

### 3.1.5 Bandwagon Effect

Independent of the COVID context, most people have a cognitive bias that influences them to behave or think the same way that they believe the crowd is [76]. Governments can make participation more visible to others, thus encouraging them to follow suit. For example, scanning QR codes when entering a venue forms a visual compliance that may be harder to find in systems that use passive methods like Bluetooth or GPS. Reporting statistics that demonstrate people are participating may also motivate others to join the crowd and avoid being left out. However, if people can see that the participation rates are low, then they may continue to go with the crowd and stay away. Alternatively, the theory of diffusion of innovation suggests that there will be innovators and early adopters, and a critical mass needs to be sustained in order to reach the majority of individuals [77], although to our knowledge this has not been validated in a DCT context.

### 3.2 Access

Can people get their hands on the necessary equipment to participate in DCT? Are the people who are motivated and willing also able to afford a smartphone? Generally, there may be higher uptake if a DCT tool is available for every motivated person to use, but there are challenges in system design and path dependencies that can make this unachievable. This category looks at the technological and physical barriers that people may face when trying to participate in DCT, separated into three main categories:

### 3.2.1 Hardware

Most jurisdictions are relying on smartphone-based systems, but people in marginalised groups may not own a smartphone. For example, the smartphone penetration rate in India is only approximately 34% of the population [9], which excludes hundreds of millions of people from participating in DCT.

There are also access challenges for those with older devices, such as incompatible operating systems. For most of 2020, the iPhone 6 (which was about 5-6 years old) did not have a new enough operating system version to be compatible with the Apple/Google Exposure Notification system [78]. People may also have older cell phones that serve their everyday needs, without Bluetooth or GPS technology embedded inside the device. These barriers exclude people by creating a monetary cost to participation by requiring individuals to purchase newer devices in order to participate.

### 3.2.2 Infrastructure

Most DCT tools will require an internet connection at some point in the process, whether that is to download an app or to communicate with government servers. While a number of decentralised tools can operate without a live internet connection (e.g. Bluetooth-based methods that store data locally on the device), they still need to check for exposure notifications from central servers on a regular basis or the user won't be notified of potential crossovers with infectious people. This issue of infrastructure access is especially prominent in poor or rural communities with low or no internet access. Even though the data requirements might be quite low (a few MB per day), it still attracts a cost.

The system infrastructure should also be available for individuals to use. For example, a venue-based QR code system requires those venues to have QR codes available for people to scan. If the codes are not scannable because they are obscured or damaged, then people may not be able to log that visit in the DCT tool [75]. Similarly, there have been suggestions that venue-based Bluetooth beacons could be used to automatically log people's visits [79], which would also require that the beacons are available and functional.

### 3.2.3 Software

There may also be software compatibility issues. For example, some Android devices (e.g. Huawei-branded smartphones) cannot access the Google Play app store due to a US trade ban [80]. Many jurisdictions distribute their DCT apps officially through the Google Play and Apple app stores, so users of incompatible devices have to go to extra efforts to find ways to download and install those apps. There have also been challenges with regional restrictions on some apps – for example, some users were told by the app store that their country's DCT app was not available on their device because the device was registered as being in another country, often because the user had previously lived in another country. This again required extra effort to reconfigure the app store to identify the location of the user correctly before they could participate in the DCT system.

There may also be challenges for users with disabilities, such as visually-impaired individuals who find it difficult to scan QR codes [81]. People with disabilities may not carry the same types of devices as others (i.e. disability assistive technologies that may not be running Android or iOS), so unless suitable software is provided for those devices, this could also exclude these people from participation. DCT tools that insufficiently consider the needs of people with disabilities will therefore lower the maximum achievable uptake rate, and may also be excluding a part of the population that is more vulnerable to Covid-19.

## 3.3 Skills

Do people know what DCT is and how it works? Do they know how to download an app or scan a QR code? Is information available in an easy-to-understand format? In this category, we discuss drivers and barriers about how difficult it can be for people to use DCT, depending on whether they have the skills or confidence to participate.

### 3.3.1 Technology Skills and Confidence

The development of DCT is relatively recent, and the terminology used is still new to many people. With a lack of familiarity comes a lack of confidence in their ability to use the DCT tool. People may not understand or know that Bluetooth needs to be turned on all of the time to be effective, or that permissions have to be enabled in multiple settings on their device. There may also be more general technical skills challenges, such as some people not knowing how to download and install an app, or not knowing how to interact with an app to turn DCT on. In New Zealand, there were reports of people taking photos of QR codes with their smartphone camera, rather than scanning the codes using the DCT app, meaning that they were not contributing information to the DCT system [82]. Some people may have challenges scanning QR codes reliably and quickly, and fear holding up other people waiting to enter a venue, meaning that they give up. Confidence can play a large role in supporting or inhibiting a person's use of technology; people can be afraid of pushing the wrong buttons, and end up choosing to do nothing than to do the wrong thing [83, 84].

Having sufficient technical knowledge can also help people understand how their efforts in participating in DCT is effective towards suppressing Covid-19, which also impacts their motivation and trust. Some interviews have revealed how people have outweighed the concerns of privacy over the potential benefits of using digital contact tracing [85], and that reasoning requires at least some understanding of technical concepts like data architectures. Otherwise, people may default to their underlying values and ideology to fill the gap, which may not align with the technical truth of how the system works. Conspiracy theories around DCT stem, in part, from a lack of understanding, or too much complexity, leading to people looking for simpler explanations. This intersects with the level of trust people put in the government and experts who are explaining how these tools work.

Interestingly, a study has found that older populations and people with compromised medical conditions are more likely to participate in DCT than younger populations [49], likely because older populations are more vulnerable to Covid-19 and have a higher perception of risk. Anecdotal evidence suggests that older people may have the motivation to use the app, but are challenged by not necessarily having the access or skills to do so effectively, while younger people may have access and skills, but lack the motivation.

### 3.3.2 Literacy Skills

As much of DCT is new to people, it has to be accompanied by strong communications plans and messaging campaigns that explain the concepts. Active participation in DCT may require people to follow instructions, which rely on the user's general literacy – this is a challenge highlighted for digital inclusion generally [57]. People who do not have sufficient literacy in the primary language of the jurisdiction they reside in may therefore find it difficult to participate. This could include migrants, speakers of indigenous languages in colonised areas, and less educated people, particularly where technical vocabulary is used to explain how the system works. Users may be presented with complicated privacy statements that are hard to understand and evaluate. These factors also affect a person's confidence in using the app, and therefore their motivation.

## 3.4 Trust

Will the system owners be able to conduct surveillance and know where I've been, or know who I've been talking to? Do I believe that the system does what its developers say it does? Trust allows people to feel confident that their participation in DCT is safe. This category looks at the relationships individuals have with the various entities associated with DCT, as well as the system itself.

### 3.4.1 Trust in Government

People form a relationship with their governments based on a variety of factors including political ideology, media reporting of political affairs, and past experiences with government agencies. That relationship informs whether individuals trust the government to handle their data sensitively and for the stated purposes only. DCT systems in general infringe on people's privacy, but this is balanced against the need to manage the public health crisis of the pandemic. Therefore, trust plays an important role in helping individuals determine whether they will accept some loss in privacy, or if they are worried that there may be more, hidden losses of privacy that have not been presented to them. Interviews in German-speaking countries early in the pandemic showed participants framed DCT tools as "governmental surveillance tools" to be approached with skepticism [86].

This threat manifested into reality in Singapore, when a Minister confirmed that police were able to access DCT records, despite previous assurances that this data would only be used for public health purposes. This then resulted in disappointment from the public, with many claiming that they deleted the DCT app from their phones [87, 88]. A few months later, it was revealed that Australian State police forces had accessed centralised QR code scan records multiple times [89]. A study found that when people have confidence in their government's policies and practices, they are more open to disclosing information for contact tracing [90]. However, it can be difficult to separate out confidence in a government's response to Covid-19, and confidence in the public sector more generally – the first has a short-term effect and is more volatile, while the second has a longer-term underlying effect.

Trust in government is a very personal value, and the factors that influence this vary significantly from one person to another. For example, ethnic minorities or indigenous peoples with personal conflicts from systemic injustices may have less trust in government [91, 92]. These relationships cannot be easily improved through advertising campaigns, and require longer-term commitments and corrections from governments. On the other hand, state surveillance has already been normalised in some jurisdictions, which lowers the impact of additional surveillance on trust [93].

### 3.4.2 Trust in Corporations

Similar to trust in government, people form relationships with the corporations that they interact with throughout their lives. Apple and Google have developed a large role in DCT through their Exposure Notification Framework that has been adopted by many jurisdictions. The track record of the big tech companies and their treatment of user data has been controversial in recent years, with calls for more regulation and protections for consumers [94, 95]. Some people have therefore viewed Apple and Google's involvement in DCT with suspicion, worried that they are harvesting data for commercial purposes or to build user profiles, regardless of the technical reality. This is despite Apple and Google viewing themselves as protectors of individual privacy against government overreach during the pandemic, setting restrictions on the use of their protocol that have been criticised by some public health agencies and governments [96, 97]. Again, this is a relationship that cannot be easily improved through generic and impersonal marketing, and governments that want apprehensive citizens to use an Apple/Google protocol face a difficult challenge.

### 3.4.3 Data Architecture

Whether a DCT system is centralised or decentralised has an impact on whether individuals trust the system itself, separate to their relationships with the system designers or owners. The architecture choice determines where data rests, particularly for individuals who do not have COVID-19 and therefore do not pose any risk to the rest of the public. Some jurisdictions, such as Germany, initially had a centralised architecture but then switched to decentralised in order to maintain user privacy and build trust with the public [98]. While a centralised approach could make it easier for health officials to act by providing more information for manual-led contact tracing, decentralised approaches can help give individuals more confidence that their data cannot be improperly used by the government because the data is not held by the government by default.

### 3.4.4 Security Concerns

This primarily relates to concerns of misuse of data by third parties, such as hackers. The digital revolution has brought with it many data breaches and security attacks, which can reduce people's confidence in the security of digital systems. With a very rapid response required for the pandemic, some may worry that best practice processes may have been poorly implemented or skipped during development. In Qatar, a security flaw in the Ehteraz app meant that sensitive personal information of over a million users placed at risk [99]. This threatens confidence, especially in populations that are less accustomed to using technology and are worried about scams and data breaches, thereby creating another barrier to participation in DCT.

### 3.5 Personal Agency

The drivers and barriers identified thus far interact in a complex way, ultimately leading to a decision to participate or not in DCT. We observe a similarity between many of these factors in a dependency on individual control. The feeling of a loss in control or agency is common during a pandemic, where the disease is widespread and individual-level actions do not seem to have much of an impact. This context influences the sense of control that may be gained when an individual believes that their choices and actions can create a change in the direction they want. Large-scale surveys in China found that a sense of control is associated with a high perceived knowledge of Covid-19 [100]. Giving individuals knowledge about DCT so that they can make their own decision around participating can help give them agency in a chaotic world. Particularly where they perceive that DCT is effective, it can help people feel that they are doing something to combat the pandemic. This approach may have benefits beyond collecting information for DCT – prior literature suggests participatory approaches may help improve morale and happiness [101, 102, 103], which can help with overall compliance with public health measures and therefore help keep the overall spread of the disease under control.

## 4. Influencing the Drivers and Barriers

As the previous section has shown, the uptake rate of DCT is not just determined by the technical design of the tool itself – there are internal and external factors that can also influence an individual's decision to participate. As DCT is a relatively new pandemic management tool, there has been limited research on how to increase uptake, with most governments relying on advertising alone. This section draws upon literature from psychology, public policy, human-computer interaction, marketing, and other related disciplines to propose ideas for policies and interventions that could increase uptake. Throughout this section, we have provided references to the relevant literature for the reader to find further detail and expertise. These suggestions are not comprehensive, and their appropriateness will depend on the context of each jurisdiction. There is no one-size-fits-all DCT approach, and there is no one-size-fits-all set of policy interventions to improve DCT uptake. The suggested interventions are summarised in Table 3.

| Behaviour | Policy |
|---|---|
| • Changing Behaviour Norms<br>• Goal Setting through Transparency | • Participation Incentives<br>• Strengthen Privacy<br>• Access Subsidies<br>• Mandates and Compulsion<br>• Policy Timing |
| **Marketing** | **Technology** |
| • Advertising<br>• Education and Information | • User-centred Design<br>• Transparency and Security |

Table 3. Suggested policy interventions for improving uptake of digital contact tracing. Icons: flaticon.com

### 4.1 Behaviour
#### 4.1.1 Changing Behaviour Norms

The limited research on DCT has thus far suggested that the strongest predictor for uptake rate lies in building habits into people's behaviours [49, 104]. By helping people change their routine to the "new normal", when DCT is adopted as a part of everyday life alongside other habits like using hand sanitiser and social distancing, motivation barriers are reduced and uptake rates naturally increase. Changing norms at a population level is non-trivial, and relies on many of the other policies and interventions mentioned in this section – but the objective of these efforts needs to point towards long-term behaviour change, not short-term spikes driven by fear or anxiety. While nudge theory has become popular in behavioural economics and public policy communities, there are limitations in its use for changing behaviours with urgency, especially in complex scenarios where nudges oversimplify the problem [105, 106].

#### 4.1.2 Goal Setting through Transparency

If the government clearly communicates targets for the uptake rate, and releases statistics on the current uptake rate, then people can see if there is any progress. This can help contribute towards the bandwagon effect by helping show that others are participating, motivating them to join the crowd. A study during the 2013 German Bundestag election showed how anxiety and enthusiasm can drive behaviour change when people can see how they are, or are not, contributing towards a shared goal [107]. Conversely, there have been examples of governments hiding participation statistics, claiming that it could risk public safety, which has led to assumptions that the participation rates are low [108]. Related to this is demonstrating the effectiveness of DCT by providing statistics and sharing use cases of where the tool has tangibly assisted contact tracing efforts and helped reduce the spread of the disease.

## 4.2 Policy
### 4.2.1 Participation Incentives

Financial rewards could be given for those participating in DCT systems in order to provide motivation. A study in Germany showed that even small incentives of a few dollars or euros can increase uptake rate, with a stronger effect than providing information alone [49]. The same study found that younger people are more responsive to incentives. The study also suggests that instead of paying cash for downloading an app, in-app credits may be an option, particularly as it may be easier to administer where ongoing use is to be encouraged, as long as the credits are redeemable at places that people are interested in. Lotteries have also been suggested to reduce the cost of providing incentives while still attracting participation.

However, incentive schemes require careful design to avoid incentivising excess mobility, where people move more than necessary, just to receive more reward. This may be counterproductive to suppression goals if there are active cases in the community, as extra mobility could mean more exposure events and therefore make containing an outbreak harder. Conversely, when there are no cases in an area the harms of incentivising mobility may be lower. Another incentive option may be through the use of discounts, such as a 5% meal discount for demonstrating participation in DCT, which provides a benefit for participation but is less likely to encourage excess movement. It is important to note that incentives need to be equitable – all people need to feel that they have an equal opportunity to earn incentives.

### 4.2.2 Strengthen Privacy

Stronger privacy measures can have a large influence on the drivers in the trust category. A study conducted across five countries showed that the main factors hindering uptake are related to trust [109]. A study in Taiwan showed that acceptance for tracking technologies increased from 75% to 90% amongst young adults if privacy measures are incorporated [39]. It is therefore critical for government to demonstrate a robust commitment to privacy. It is important to note that people's trust or mistrust tends to relate to their experiences with the public sector as a whole, rather than with specific agencies or departments. Beyond assurances from senior officials, privacy can also be strengthened through legislative protections for DCT data. In Australia, penalties have been introduced for misuse of DCT data, also prohibiting employers from coercing their employees to share DCT data, and requiring that data be held onshore [110]. DCT systems can also be re-designed to collect the minimal amount of necessary and relevant data, and ensure that data will be deleted at a specified time or condition in the future when it is no longer needed.

### 4.2.3 Access Subsidies

Governments can reduce barriers to accessing the necessary hardware and infrastructure for DCT by providing monetary subsidies. Subsidies could be 100%, making the device free – wearable devices such as the TraceTogether Tokens used in Singapore have been distributed in this way [111]. However, if the government issues its own devices, then parts of the population may view it with suspicion if they cannot inspect the device or know for sure how it works. Making the design open and transparent, or allowing civil society technologists to inspect the design, can assist with building trust. Alternatively, people can be given choices to acquire devices from commercial vendors. For example, a phone discount voucher could allow individuals to purchase a new device, while giving them the flexibility to decide what device would be most suitable for their needs and budget.

For DCT systems that require an active internet connection, zero-rating is a technical tool that allows specific applications to have free mobile internet access [112]. In South Africa, zero-rating has been used to help those without internet access participate, although there have been some issues in the downloading process [113]. Improving access to the necessary hardware and infrastructure can help improve uptake rates for DCT.

### 4.2.4 Mandates and Compulsion

As seen in jurisdictions like Qatar, China, and Singapore, mandatory participation in DCT can strongly motivate people and lead to a high uptake rate. However, mandates to use DCT are often practically unenforceable, because the system runs autonomously in the background of a smartphone and it would be time consuming and resource intensive for businesses or police to check. DCT systems that have very visible components, such as showing a wearable device or showing a QR code scan log, may be easier to enforce. A quasi-mandate (e.g. mandatory but unenforced) may be better than a purely voluntary approach, as this sends a clear signal and can help drive behaviour change. However, this may still attract criticism from libertarian ideologues opposed to such overt government behaviour manipulation. Large businesses and venues (e.g. supermarkets, concerts and festivals) with higher risk levels could be required to assign staff at entrances to strongly encourage people to demonstrate use of DCT before entering premises, with leniency and empathy for those who cannot participate.

A mandate without consideration for the digitally excluded may lead to more problems for the government. In India, as the Aarogya Setu app became compulsory, the majority of the population did not have access to smartphones and could not access the app [114]. This has led to criticism that the DCT tool is not effective and contributed towards reduced trust in government. A mandate can generate a backlash in public opinion from people who cannot participate, who feel anxious about not being able to protect themselves, and feel left out by their government. A further challenge for governments is that those using the Apple/Google Exposure Notification Framework are unable to make use of their DCT tool mandatory, as Apple and Google's rules prohibit compulsion and they have indicated that they will disable access to the framework within infringing jurisdictions [115].

### 4.2.5 Policy Timing

As identified in 3.1.2, changes in perception of risk can be a strong motivator for people to take action to protect themselves, potentially driving them towards participation in DCT. Therefore, if a government has policy changes, it may be useful to time those interventions around changes in risk levels, taking advantage of heightened motivation to mitigate risk – this has been discussed in other contexts such as climate change [116, 117]. For example, changes could be introduced at a time when populations are coming out of lockdown, or when more infectious strains of the virus have developed, so that they are front of mind for individuals. However, governments may be criticised for sitting on good ideas and policies and not implementing them earlier, so we view this as a weaker option for policymakers.

### 4.3 Marketing
### 4.3.1 Advertising

Most governments have developed strong communications plans to provide information to their populations about Covid-19, and integrating messaging about DCT is important for developing awareness. In advertising the use of DCT, one recommendation is to focus on making it relatable to people [118]. For example, highlighting stories of how DCT can or has helped everyday people in the community may be more effective than simply telling people to "download the app" or "turn Bluetooth on". Some studies have found that relatable media, particularly on social media, can have a stronger influence on younger populations [119, 120]. Advertisements by influencers or well-known experts can also help generate more empathy to influence uptake [121]. In Taiwan, official messaging embraced the use of humour and memes, aiming to make true information more easily shareable than misinformation [122]. The literature is less clear on the value of collectivist messaging, and its suitability likely depends on the context of each culture and jurisdiction [123, 124, 125]. Emotional messaging, both prosocial and threatening, has been shown to be effective at changing behaviours in the context of Covid-19 [126, 127]. Governments should be aware of the limitations of one-to-many advertising, in that it favours short messages that can make it difficult to explain nuance or build a relationship with the public. Using a variety of messaging pushes targeted at different communities, particularly more vulnerable or marginalised groups, can help inform all people and avoid exclusion.

### 4.3.2 Education and Providing Information

Beyond advertising, the government can fund initiatives that push information more comprehensively. Going into communities to help train people for DCT participation can be beneficial for increasing uptake, which can build upon existing systems for digital literacy training [121]. Rather than posting TraceTogether Tokens to individuals, Singapore set up collection points so that staff could explain how the technology worked and what exactly people needed to do to use it [128]. Anchoring DCT in the Covid-19 context can also be important for ensuring that individuals appropriately perceive the level of risk and the necessity to participate in such training [129].

Governments also have to fight misinformation and disinformation, particularly on social media platforms. The privacy perception of an individual matters more in their decision-making process than the reality. Providing a compilation of verified information through official channels can provide a reference point for individuals seeking more information. However, waiting for individuals to find this information is insufficient, and governments may need to be more proactive and pre-bunk predictable ideas that may arise – there is significant research in vaccine hesitancy and conspiracy theories that explores the same themes [130].

## 4.4 Technology
### 4.4.1 User-centred Design

Generally, a well-designed tool with good usability is more likely to be adopted by individuals than tools that are designed without user experience in mind [131]. Users are increasingly prioritising convenience in their decision making, and unnecessary skills barriers will harm uptake [132, 133]. User testing is an important process for ensuring that the bias of system developers and designers does not lead to unintended consequences with real users. Software projects, even when delivered successfully and on time, can fail if the users are dissatisfied [134]. User experience has been an under-discussed area of DCT, even though its impact is evident through App Store reviews [54].

Software design is often an iterative process, and so if improvements are made to DCT tools after launch, these changes should also be clearly communicated to the public who may have had poor experiences with the tool in the past and given up. It is critical to improve the DCT system to incorporate best practice seen in other jurisdictions, and adapt as we learn more about the disease. Ensuring that DCT tools are designed in a way that is available to as wide an audience as possible is also important. This may include supporting multiple languages, compatibility for assistive technology devices, or including haptic feedback for the visually-impaired and/or hearing-impaired.

There have also been suggestions of improving motivation by gamifying DCT tools, particularly those that require active participation (e.g. scanning QR codes at venues) [42, 135]. There could be non-financial rewards like congratulatory messages, badges and achievements, and celebrating streaks for participating on consecutive days. However, gamification carries many of the same risks as the financial incentives described in 4.2.1, and has to be carefully designed to avoid incentivising excess mobility. There may also be concerns that providing psychological incentive can "crowd out" moral motivations for participation in DCT [136]. Simply providing feedback to users that the system is working as intended, and that data is being logged correctly or that matches are being made, can help influence the perception that the DCT system is effective and therefore worth continuing to participate in.

There have also been suggestions that systems that require active participation (e.g. scanning QR codes or confirming ongoing participation in the DCT system) could use period reminders through smartphone notifications to prompt users to participate. However, these should also be considered carefully as having too many notifications can lead to fatigue, where the messages are ignored by users (or even blocked from appearing).

### 4.4.2 Transparency and Security

Beyond the choice of centralised or decentralised data architectures, providing transparency on system design can be helpful for providing confidence to individuals that the DCT system does what the government claims it does, building trust. Making the code open-source means that it is released online to be publicly accessible, where it can be inspected by others and potentially improved upon by the community as well [137]. Independent security and privacy audits should also be conducted by trusted entities [138], and providing better data sovereignty could also improve confidence and therefore trust [139, 140]. Governments should acknowledge the limitations of the technology, and give confidence that appropriate oversight and accountability mechanisms are in place.

### 5. Conclusions

Digital contact tracing has a role to play in responding to infectious disease pandemics like COVID-19. The participation or uptake rate is a key determinant in the effectiveness of these tools. As a new tool that has seen very limited use prior to this pandemic, there is limited published literature about how to use DCT, or the most effective ways to encourage participation, although more work is being published every day and this paper only represents a point in time. By exploring how DCT has been implemented in practice, and seeing the variety of different technologies and approaches in each jurisdiction, we can now start to see some trends and patterns for successful and unsuccessful introduction of DCT. Through a literature review approach, this paper has used a Motivation, Access, Skills, and Trust framework to help explore the drivers and barriers to uptake for DCT tools. We also look into tangible policy actions and interventions that government could take to influence these drivers and barriers, borrowing from the literature in related areas. More experimental research into effective policy design is required to build an empirical evidence base for DCT. Measuring perceptions of demographic groups and marginalised communities may also help with understanding how the drivers and barriers differ between those groups. We hope that this work may help policymakers with their thinking around how to build engagement and participation with the public in DCT, both for Covid-19 and for future pandemics. The use of the MAST framework in this paper indicates that there may also be broader opportunities for learning from the deployment of DCT for policy efforts towards digital inclusion.

**Dr. Andrew Tzer-Yeu Chen** is a Research Fellow with Koi Tū: The Centre for Informed Futures, a transdisciplinary think-tank at The University of Auckland, New Zealand. His background is in computer engineering, investigating computer vision surveillance and privacy. His research interests now sit at the intersection of digital technologies and society.

**Kimberly Thio** was a summer research student with Koi Tū: The Centre for Informed Futures. She is studying psychology and global studies at The University of Auckland, New Zealand.